\documentclass[aps,pre,twocolumn,showpacs,
amsmath,amssymb,reprint]{revtex4-1}

\usepackage{graphicx}
\usepackage{dcolumn}
\usepackage{bm}

\begin{document}

\preprint{APS/123-QED}

\title{Hard-Sphere Brownian Motion in Ideal Gas\,: \\
Inter-Particle Correlations, Boltzmann-Grad Limit,\\
and Destroying the Myth of Molecular Chaos Propagation}

\author{Yuriy E. Kuzovlev}
\affiliation{Donetsk A.A.Galkin Institute
for Physics and Technology of NASU,\\
ul.\,R.\,Luxemburg 72, Donetsk 83114, Ukraine}
\email{kuzovlev@fti.dn.ua}

\date{10 January 2010}

\begin{abstract}
The BBGKY hierarchy of equations for a particle interacting with
ideal gas is analyzed in terms of irreducible many-particle
correlations between gas atoms and the particle's motion. The
transition to the hard-sphere interaction is formulated from
viewpoint of the recently discovered exact relations connecting the
correlations with the particle's probability distribution. Then the
Boltzmann-Grad limit is considered and shown not to lead to the
Bolzmann hierarchy and the molecular chaos, since correlations of all
orders keep significant.in this limit, merely taking a singular form.
\end{abstract}



\maketitle


\section{\,Introduction}

In the work \cite{tmf}, as well as in its arXive preprints
\cite{artmf} and premising works \cite{pro,mpa3}, based on principles
and methods of rigorous statistical mechanics, - first of all, on the
Bogolyubov approach to it \cite{bog}, - I proved existence of general
exact relations (``virial relations'') connecting (i) probability law
of random walk of a test ``Brownian'' particle (BP) in a fluid and
(ii) irreducible many-particle statistical correlations between
molecules of the fluid and the BP's walk (in particular, BP may be
merely a marked molecule). I emphasized also that these relations
make it clearly visible that all $\,n$--particle correlations always
are quantitatively significant, even under the low-density
(``Boltzmann-Grad'') gas limit. This fact, in turn, implies that the
Boltzmannian kinetics is incorrect even in this limit, and true
kinetics of fluids, - including low-density gases! - should take into
account all the correlations.

The meaning of so universal correlations was explained already in
\cite{i1}\, (this article is hardly available on-line, but one can
read its my own author's translation \cite{bbgky} from Russian
original, or see preprint \cite{i2}). It is sufficient to notice that
a fluid as the whole (and first of all dilute gas!) is indifferent to
a number of past collisions happened to its given particular
molecule. Therefore any molecule has no definite (\,{\it a priori}\,
predictable) ``time-average'' rate of collisions \cite{ex,kr}. In
other words, actual rate of its collisions undergoes fluctuations
which are indifferent to time averaging. Thus that are scaleless
fluctuations with 1/f type spectrum \cite{com}. The mentioned statistical
correlations directly reflect complicity of particles (via mutual
collisions) in these fluctuations and, hence,  indirectly describe
their statistics (for detail see \cite{i1} or \cite{bbgky} or
\cite{i2}).

As far as I know, first such statements about molecular
motion were put forward in works by G.Bochkov and me
\cite{pr1,bk1,bk2,pr2,bk3,pjtf} as conjectures about
origin and properties of 1/f noise accompanying charge transport
(i.e. Brownian motion of charge carriers) and other
transport processes (in generalized sense \cite{i2}).
We demonstrated once again that correct ideology
leads to useful results even at phenomenological level.  In particular,
in \cite{bk2,pr2} it was shown that fluctuations of rate of collisions
(or, equivalently, 1/f fluctuations of diffusivity and mobility) possess
essentially non-Gaussian statistics gravitating towards power-law
probability tails.

The latter circumstance requires, in view of the well known
Marcinkiewicz theorem (see e.g. \cite{luc}), to deal with whole
infinite chain of cumulants of the fluctuations. On the other  hand,
according to the later works made at the microscopic level, firstly
\cite{i1} and then its development in \cite{p12}, $\,n$-order
cumulant  of fluctuations in the rate of collisions associates with
specifically  $\,(n+1)$-particle correlations (irreducible component
of $\,(n+1)$-particle distribution function). That is why one should
not truncate the BBGKY hierarchy of equations!  Anyway, neglect of
three-particle (and thus higher) correlations means rejection of the
fluctuations at all (like neglect of two-particle correlations
rejects any collisions at all \cite{rev}).

The aforesaid was confirmed by exact results of \cite{tmf,artmf,pro}.
Nevertheless, a specially visual analysis of inter-particle
correlations may be useful. This is just one of purposes of the
present paper.

With this purpose it is natural to concentrate on the Brownian motion
of (molecular-size) particle in ideal gas (see \cite{igp} or some of
preprints \cite{ig} and also \cite{tmf,artmf}), which is most simple
``kinetic process'' since it produces least amount of inter-particle
correlations. Besides, this is good motive to scan the limit
transition from smooth interaction (between BP and gas atoms) to
singular ``hard-sphere'' one dividing into momentary ``collisions''.

Another our purpose is to perceive falsity of popular treatments of
the ``hard-sphere BBGKY hierarchy of equations''
\cite{gr,lan,ch,re,bnn,vblls,sp1,gp1,gp2,sp2} trying to reduce it, in
the Boltzmann-Grad limit, to so-called ``Boltzmann hierarchy''.

We will begin in Sec.2 by formulation of the BBGKY hierarchy for our
particular problem and corresponding (above mentioned) exact virial
relations \cite{tmf,artmf,igp,ig}. After their consideration in Sec.3
we will see for ourselves that collisions constantly give birth to
various statistical correlations, and not only post-collision ones
but also pre-collision correlations. Then, in Sec.4, discuss and
execute the hard-sphere limit of our BBGKY hierarchy, taking in mind
that it should stay compatible with the virial relations as they are
independent on character of interactions (interaction potentials).
The result is (i) modified BBGKY equations combined with (ii)
familiar boundary conditions to them at hyper-surfaces (in
many-particle phase spaces) corresponding to collisions. The second
ingredient allows to transform the first into usual ``hard-sphere
BBGKY hierarchy'' but, at the same time, it forbids its reduction to
the  ``Boltzmann hierarchy'' \cite{lan,ch,vblls,sp1,gp2} under the
Boltzmann-Grad limit. Thus, the correct theory does not present such
marvellous simplifications as ``propagation of chaos'' and closed
equation for one-particle distribution function (DF): as before, one
has to solve an infinite hierarchy of equations! This situation will
be discussed in Sec.5.

\section{BBGKY hierarchy, its cumulant representation,
and virial relations}
We want to consider random walk $\,{\bf R}(t)\,$ of a Brownian
particle (BP) in thermodynamically equilibrium ideal gas, assuming
that at initial time moment $\,t=0\,$ BP starts from certainly known
position $\,{\bf R}(0)=0\,$. The BBGKY equations for this problem can
be either derived \cite{ig} directly from the Liouville equation,
following Bogolyubov \cite{bog}, or extracted \cite{igp} from general
results of \cite{tmf,artmf,pro}. They reads as
\begin{equation}
\frac {\partial F_k}{\partial t}=[\,H_{k},F_k\,]+n \,\frac {\partial
}{\partial {\bf P}}\int_{k+1}\!\!\Phi^{\,\prime}({\bf R}-{\bf
r}_{k+1}) \,F_{k+1}\,\,\,,\label{fn}
\end{equation}
where $\,k=0,1, ...\,$,\, $\,H_{k}\,$ is Hamiltonian of subsystem
``\,$k$ atoms + BP\,'',\, $\,[...,...]\,$ means the Poisson
brackets,\, $\,\Phi({\bf r})\,$ is potential of (short-range
repulsive) interaction between BP and atoms,\,
$\,\Phi^{\,\prime}({\bf r}) =\nabla\Phi({\bf r})\,$,\,  $\,n\,$ is
gas density (mean concentration of atoms),\, $\,F_0=F_0(t,{\bf
R},{\bf P}\,;n\,)\,$\, is normalized DF of the BP's coordinate
$\,{\bf R}\,$ and momentum $\,{\bf P}\,$,\, $\,F_k=F_k(t,{\bf R},
{\bf r}^{(k)},{\bf P},{\bf p}^{(k)}\,;n\,)\,$\, are
$\,(k+1)$-particle DFs for BP and $\,k\,$ atoms \cite{comm1},\,
$\,{\bf r}_j\,$ and $\,{\bf p}_j\,$ denote coordinates and momenta of
atoms, respectively,\, $\,{\bf r}^{(k)} =\{{\bf r}_1...\,{\bf
r}_k\,\}\,$,\, $\,{\bf p}^{(k)} =\{{\bf p}_1...\,{\bf p}_k\,\}\,$,\,
and\, $\,\int_k ... =\int\int ...\,\,d{\bf r}_k\,d{\bf p}_k\,$.

Initial conditions to these equations, corresponding to the gas
equilibrium, at temperature $\,T\,$, are
\begin{equation}
\begin{array}{c}
F_k|_{t=\,0}\,=\delta({\bf R})\, \exp{(-H_k/T\,)}= \label{ic}\\
= \delta({\bf R})\,G_M({\bf P}) \prod_{j\,=1}^k E({\bf r}_j-{\bf
R})\, G_m({\bf p}_j)\,\,,
\end{array}
\end{equation}
where\, $\,M\,$ and $\,m\,$ are masses of BP and atoms,
respectively,\, $\,G_m({\bf p})=(2\pi Tm)^{-3/2}\exp{(-{\bf
p}^2/2Tm)}\,$ is the Maxwell momentum distribution of a particle with
mass $\,m\,$,\, and $\,E({\bf r})=\exp{[-\Phi({\bf r})/T\,]}\,$.
Besides, existence of the thermodynamical limit presumes the
``cluster property'' of DFs. that is vanishing of inter-particle
correlations under large spatial separation of particles, so that\,
$\,F_k\,\rightarrow\,F_{k-1}\,G_m({\bf p}_s)\,$\, at \,$\,|{\bf r}_s
-{\bf R}|\rightarrow\infty\,$\,, where $\,1\leq s\leq k\,$ and
$\,F_{k-1}\,$ does not include $\,{\bf r}_s\,$ and $\,{\bf p}_s\,$.
That are boundary conditions to Eq.\ref{fn}.

In view of these conditions, in order to visually extract
inter-particle correlations, it is convenient to make the linear
change of variables, from the DFs $\,F_k\,$ to new functions
$\,C_k\,$,\, \cite{comm2} as follow:
\begin{equation}
\begin{array}{c}
F_0(t,{\bf R},{\bf P};n)\,= \,C_0(t,{\bf R},{\bf P};n)\,\,\,,\\ \\
F_1(t,{\bf R}, {\bf r}_1,{\bf P},{\bf p}_1;n)\,=\,  \\
=\,C_0(t,{\bf R},{\bf P};n) \,f({\bf r}_1\!-{\bf R},{\bf p}_1)\,+\\
+\, C_1(t,{\bf R}, {\bf r}_1,{\bf P},{\bf p}_1;n)\,\,\,,\\ \\
F_2(t,{\bf R},{\bf r}^{(2)},{\bf P},{\bf p}^{(2)};n)\, = \\
=\,C_0(t,{\bf R},{\bf P};n) \,f({\bf r}_1\!-{\bf R},{\bf
p}_1)\,f({\bf r}_2\!-{\bf R},{\bf p}_2)\,+\\+\,C_1(t,{\bf R},{\bf
r}_1,{\bf P},{\bf p}_1;n) \,f({\bf r}_2\!-{\bf R},
{\bf p}_2) +\\
+ \,C_1(t,{\bf R},{\bf r}_2,{\bf P},{\bf p}_2;n)
\,f({\bf r}_1\!-{\bf R}\!,{\bf p}_1)\,+ \\
+ \,C_2(t,{\bf R},{\bf r}^{(2)}, {\bf P},{\bf
p}^{(2)};n)\,\,,\label{cf}
\end{array}
\end{equation}
and so on.\, with\, $\,f({\bf r},{\bf p}) = E({\bf r})\,G_m({\bf
p})\,$.

Clearly, such defined $\, C_k\,$ can be named cumulant functions (CF)
since represent irreducible correlations between $\,k\,$ gas atoms
and total BP's path $\,{\bf R}\,$. In their terms the BBGKY hierarchy
(\ref{fn}) takes the form \cite{tmf,artmf,ig}
\begin{eqnarray}
\frac {\partial C_{k}}{\partial t}=[H_k,C_k]+n \,\frac {\partial
}{\partial {\bf P}}\int_{k+1}\!\! \Phi^{\,\prime}({\bf
R}-{\bf r}_{k+1})C_{k+1}+\nonumber\\
+\,T\sum_{j\,=1}^{k}\, G_m({\bf p}_j) \,E^{\prime}(\rho_j)
\left[\frac {{\bf P}}{MT}+\frac {\partial }{\partial {\bf P}}\right ]
\,\mathrm{P}_{kj}\,C_{k-1}\,\,, \,\,\,\,\,\label{vn}
\end{eqnarray}
where\, $\,E^{\prime}({\bf r})=\nabla E({\bf r})\,$,\,
$\,\rho_k\,\equiv\, {\bf r}_k\!-{\bf R}\,$,\, and\,
$\,\mathrm{P}_{kj}\,$\, means replacement of pair of arguments
$\,x_j=\{{\bf r}_j,{\bf p}_j\}\,$\, (if it is present) by
$\,x_k=\{{\bf r}_k, {\bf p}_k\}\,$. The mentioned initial and
boundary conditions simplify to
\begin{equation}
\begin{array}{c}
C_0(0\,,{\bf R},{\bf P};\,n)\,=\delta({\bf R}) \,G_M({\bf P})\,\,,\\
C_{k}(0\,,{\bf R}, {\bf r}^{(k)},{\bf P},{\bf
p}^{(k)};n)=0\,\,,\label{icv}\\ C_k(t,{\bf R}, {\bf r}^{(k)},{\bf
P},{\bf p}^{(k)};n)\rightarrow0\,\,\,\, \,\,\,\texttt{at}\,\,
\,\,\,\, {\bf r}_j -{\bf R}\rightarrow \infty
\end{array}
\end{equation}
($\,1\leq j\leq k\,$). A careful enough scanning of these equations
results in observation \cite{ig} that exact relations
\begin{eqnarray}
\frac {\partial }{\partial n}\,\,
C_{k}(t,{\bf R}, {\bf r}^{(k)},{\bf
P},{\bf p}^{(k)};n)\,=\,\, \,\,\,\,  \,\,\,\,\,\,\,\,
\,\,\,\, \,\,\, \,\,\,\,\,\,\, \label{me}\\
=\,\int_{k+1} C_{k+1}(t,{\bf R}, {\bf r}^{(k+1)},{\bf P},{\bf
p}^{(k+1)};n)\,\,\,.\nonumber
\end{eqnarray}
take place. This is particular case of general ``virial relations''
found in \cite{tmf,artmf,igp} as exact properties of solutions to
BBGKY equations describing molecular Brownian motion in fluids (they
can be also deduced \cite{pro} from the generalized
fluctuation-dissipation relations \cite{fds,p}).

These relations demonstrate existence and significance of all
many-particle correlations. To see straight away that they keep
significant in the Boltzmann-Grad limit (BGL), let us introduce
characteristic interaction radius $\,r_0\,$ of the potential
$\,\Phi(\rho )\,$, corresponding free-path length of BP,
$\,\lambda=(\pi r_0^2 n)^{-1}\,$, and integrated correlations
\[
W_k(t,{\bf R};\lambda)\,\equiv \,n^k \! \int \left[\int_1\!\!
...\!\int_k
 C_{k}(t,{\bf R}, {\bf r}^{(k)},{\bf P},{\bf
p}^{(k)};n)\,\right] d{\bf P}
\]
Thus, $\,W_0(t,{\bf R};\lambda)= \int F_0(t,{\bf R},{\bf P};1/\pi
r_0^2\lambda)\,d{\bf P}\,$ is probability distribution of the BP's
path. Then Eqs.\ref{me} yield
\begin{eqnarray}
W_k(t,{\bf R};\lambda) = \frac {1}{\lambda^{k}}\! \left(\frac
{\partial }{\partial \lambda^{-1}}\right)^k W_0(t,{\bf R};\lambda)
 \,\label{intc}
\end{eqnarray}
These exact relations hold at any values of $\,r_0\,$ and $\,n\,$,
hence, under BGL ($\,r_0\rightarrow 0\,$, $\,n\rightarrow\infty\,$,
$\,\lambda =\,$const) too. They show that anyway by order of
magnitude $\,W_k(t,{\bf R};\lambda)\sim W_0(t,{\bf R};\lambda)\,$.

\section{Pre-collision correlations and failure of
molecular chaos}

Is the Boltzmann's molecular chaos (``Sto{\ss}ahlansatz'') compatible
with virial relations (\ref{me}), (\ref{intc})? May be, correlations
there are purely post-collision and therefore do not contradict
Boltzmann s ansatz proclaiming absence of pre-collision correlations?

Unfortunately, this is vain hope, and the answer is no. To make sure
of this, let us agree that post-collision ({\it out\,}-) and
pre-collision ({\it in\,}-) states of BP and an atom satisfy
$\,(\rho\cdot {\bf u})>0\,$ or $\,(\rho\cdot {\bf u})<0\,$,
respectively, - where $\,\rho={\bf r}-{\bf R}\,$, $\,{\bf u}={\bf
v}-{\bf V}\,$, with $\,{\bf V}={\bf P}/M\,$ and $\,{\bf v}_j={\bf
p}_j/m\,$ being velocities, - and consider the first three of
equations (\ref{vn}),
\begin{eqnarray}
\frac {\partial C_0}{\partial t}=-{\bf V}\!\cdot\! \frac {\partial
C_0}{\partial {\bf R}}- n \,\frac {\partial }{\partial {\bf
P}}\int_1\! \Phi^{\,\prime}(\rho_1)\,C_{1} \,\,\,, \,\,\, \label{vn0}
\,\,\,\, \,\,\,\,  \,\,\,\, \\
\frac {\partial C_1}{\partial t}=-{\bf V}\!\cdot\! \frac {\partial
C_{1}}{\partial {\bf R}}+ \mathbf{L}_1\,C_1 - n \,\frac {\partial
}{\partial {\bf P}}\int_{3}\! \Phi^{\,\prime}(\rho_2)\,C_{2} \,
+\nonumber\\ +\,T\,  G_m({\bf p}_1) \,E^{\prime}(\rho_1) \left[\frac
{{\bf P}}{MT}+\frac {\partial }{\partial {\bf P}}\right ]
C_{0}\,\,\,\,, \,\,\,\,\,\,\,\, \,\, \label{vn1} \\
\frac {\partial C_2}{\partial t}=-{\bf V}\!\cdot\! \frac {\partial
C_2}{\partial {\bf R}}+ (\mathbf{L}_1 +\mathbf{L}_2)\,C_2 - n \,\frac
{\partial }{\partial {\bf P}}\int_{3}\!
\Phi^{\,\prime}(\rho_3)\,C_{3} \, +\nonumber\\
+\,T\,(1+\mathbf{P}_{21})\, G_m({\bf p}_2) \,E^{\prime}(\rho_2)
\left[\frac {{\bf P}}{MT}+\frac {\partial }{\partial {\bf P}}\right ]
C_{1}\,\,\,\, \, \,\,\,\,\, \,\, \, \,\,\, \label{vn2}
\end{eqnarray}
Here, we introduced the Liouville operator
\[
\mathbf{L}_j = ({\bf V}-{\bf v}_j)\!\cdot\! \frac {\partial
}{\partial \rho_j} + \Phi^{\,\prime}(\rho_j)\!\cdot\! \left(\frac
{\partial }{\partial {\bf p}_j} - \frac {\partial }{\partial {\bf
P}}\right)\,\,
\]
and made use of the relative atoms' coordinates $\,\rho_j\,$.

Evidently, the last term in Eq.\ref{vn1} represents generation of
pair correlation by BP-atom interaction, and the result is
``post-collision correlation'' since it passes to finishing ``{\it
out\,}-state'' of the particles.  If we neglected three-particle
correlation represented by CF $\,C_2\,$ then solution to Eq.\ref{vn1}
would be just this post-collision correlation: $\,C_1=C_1^{out}\,$.
Substitution of this $\,C_1\,$ to Eq.\ref{vn0} would yield a closed
``Boltzmann-Lorentz'' equation for the BP's distribution
$\,C_0=F_0\,$, thus realizing the dream of ``molecular chaos''.

However, the last term in Eq.\ref{vn2}, for $\,C_2\,$, quite
similarly (even under neglect of $\,C_3\,$) produces three-particle
correlations out of the two-particle one, $\,C_1^{out}\,$. Resulting
$\,C_2\,$ contributes, via integral in Eq.\ref{vn1}, back to
evolution of $\,C_1\,$, and produces, in particular, pre-collision
pair correlations between particles going to meet one another. In
such way, $\,C_1\,$ acquires pre-collision component $\,C_1^{in}\,$,
thus turning the dream into a rough approximation. This is
illustrated schematically by Fig.1.

\begin{figure}
\includegraphics[width=8cm]{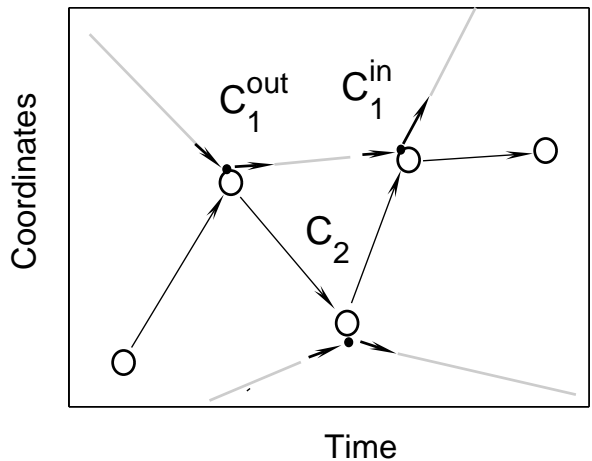}
\caption{\, A simplest diagram transforming post-collision pair
correlation, $\,C_1^{out}\,$, into pre-collision pair correlation,
$\,C_1^{in}\,$, through three-particle correlation, $\,C_2\,$. The
``Brownian particle'' is represented by circles and entire arrows
while gas atoms by dots, light lines and short arrows.}
\end{figure}

\section{Hard-sphere limit}
For further, first notice that Eqs.\ref{vn}, \ref{vn0}-\ref{vn2} can
be written in equivalent form
\begin{eqnarray}
\frac {\partial C_{k}}{\partial t}= -{\bf V}\!\cdot\! \frac {\partial
C_{k}}{\partial {\bf R}}-n \,\frac {\partial }{\partial {\bf
P}}\int_{k+1}\!\! \Phi^{\,\prime}(\rho_{k+1}) \,C_{k+1}\,+\nonumber\\
+\sum_{j\,=1}^{k}\,\mathbf{L}_j \,\, [\,C_{k}+ G_m({\bf p}_j)
\,E(\rho_j)\,\mathrm{P}_{kj}\, C_{k-1}]\,\, \,\,\,\,
\,\,\,\label{vnn}
\end{eqnarray}
Since these equations, as well as virial relations (\ref{me}) and
(\ref{intc}), are valid at any smooth interaction potential, we can
extend them to the limit of ``hard-sphere'' interaction, for
instance, defined by \,$\,\Phi({\bf r})=(r_0/|{\bf r}| )^p\,\,
\Phi_0\,$ with\, $\,p\rightarrow \infty\,$ and $\,\Phi_0 =\,$const
$\sim T\,$. At that, clearly, the BP's free path length must tend to
a constant, $\,\lambda \rightarrow \,$const\,. Therefore at any stage
of this limit transition all CFs $\,C_k\,$ are equally smooth
functions of time (except the very beginning of evolution), momenta
and coordinates, excluding ``collision regions''\, $\, |\rho_j|< r_0
+\epsilon\,$, where $\,\epsilon \rightarrow 0\,$ along with
characteristic thickness h of the potential wall\, $\,\delta = r_0/p
\rightarrow 0\,$ (at that, satisfying
$\,\epsilon/\delta\rightarrow\infty\,$).

In these collision regions, in opposite, $\,C_k\,$ ($\,k>0\,$) become
more and more sharp functions of $\,\rho_j\,$\,, so that\,
$\,\partial C_k/\partial \rho_j \sim \delta^{-1} C_k\,$, similarly to
$\, \Phi^{\,\prime}(\rho_{j}) \sim \delta^{-1} \Phi(\rho_{j})\,$\,.
Hence, the second and the fourth right-hand terms in Eq.(\ref{vn1}),
Eq.(\ref{vn2}), etc., both are infinitely increasing $\,\propto
\delta^{-1} \,$ in collision regions and, in view of absence of other
such terms, should compensate one another. Equivalently, the
expressions in square brackets in Eqs.\ref{vnn} should become
eigenfunctions of operators $\,\mathbf{L}_j \,$ corresponding to zero
eigenvalues, that is
\begin{eqnarray}
\mathbf{L}_j \,\, [\,C_{k}+ G_m({\bf p}_j)
\,E(\rho_j)\,\mathrm{P}_{kj}\, C_{k-1}]\,=\,0\, \,\,\, \,\,\,
\label{cr1}
\end{eqnarray}
asymptotically, inside the collision regions.

This statement. in its turn, means that expression in the square
brackets represents integral of motion, and its values at beginning
and at end of the pair collision,are equal. Thus, visualizing only
variables what change during collision, we can write
\begin{eqnarray}
C_{k}({\bf P}^{\,in},{\bf p}^{\,in}_j)\,+ G_m({\bf
p}^{\,in}_j)\,\mathrm{P}_{kj}\,  C_{k-1}({\bf P}^{\,in})
\,= \,\,\, \,\,\,\,\, \,\,\,\, \label{cr2}\\
=\,C_{k}({\bf P}^{\,out},{\bf p}^{\,out}_j)\,+ G_m({\bf
p}^{\,out}_j)\,\mathrm{P}_{kj}\,  C_{k-1}({\bf P}^{\,out}) \,\,,
\,\,\,\,\,\, \nonumber
\end{eqnarray}
where\, $\,\rho_j=r_0\Omega\,$, with $\,\Omega\,$ being unit normal
vector, and\, {\it in\,}-states and {\it out\,}-states correspond to
$\,\Omega\cdot {\bf u}_j<0\,$ and $\,\Omega\cdot {\bf u}_j>0\,$,
respectively. and are connected via the limit collision dynamics
(mirror reflection):
\begin{eqnarray}
{\bf P}^{\,in}+{\bf p}^{\,in}_j\,= \,{\bf P}^{\,out}+{\bf
p}^{\,out}_j\,\,\,, \,\,\,\, \,\,\,\, \,\,\,\,\,
\,\,\,\, \,\,\,\, \,\,\,\, \,\,\,\,\,\,\, \nonumber\\
\Omega\cdot({\bf v}^{\,out}_j-{\bf V}^{\,out})\, =\,-\,
\Omega\cdot({\bf v}^{\,in}_j-{\bf V}^{\,in})\,\,\,,
 \,\, \,\,\,\, \,\,\,\, \,\,\,\,\, \,\,\,\label{mr}\\
(\,1-\Omega \otimes \Omega )\,({\bf v}^{\,out}_j-{\bf V}^{\,out})\,
=\, (\,1-\Omega \otimes \Omega )\,({\bf v}^{\,in}_j-{\bf
V}^{\,in})\,\,\nonumber
\end{eqnarray}

On the other hand, integration of (\ref{cr1}) over the collision
region, - for instance, at $\,j=k\,$, -  yields
\begin{eqnarray}
-\frac {\partial }{\partial {\bf P}}\int_{k}
\Phi^{\,\prime}(\rho_{k}) \,C_{k}\,=\,-\,\widehat{\Gamma }\,
C_k\,\equiv \,\,\,\,\,  \,\,\,\, \,\,\,\, \,\,\,\,\, \,\,\,\,\,
\,\,\,\,\, \,\,\,\,\, \,\,\,\,\,
\,\,\,\,\, \,\,\,\,\, \label{cr3}\\
\,\,\,\,\, \,\,\,\,\,\equiv  \,r_0^2 \int \int (({\bf v}_k -{\bf
V})\cdot\Omega )\,\, C_k(\rho_k=r_0\Omega )\,\, d{\bf p}_k\,d\Omega
\,\, \nonumber
\end{eqnarray}

Substituting this equality into Eqs.\ref{vnn}, for complement of the
now forbidden regions $\,|\rho_j|\leq r_0\,$ we obtain
\begin{eqnarray}
\frac {\partial C_{k}}{\partial t}= -{\bf V}\!\cdot\! \frac {\partial
C_{k}}{\partial {\bf R}}-\sum_{j\,=1}^{k}\,{\bf u}_j \cdot\! \frac
{\partial C_{k}}{\partial \rho_j} \, -\,n \,\widehat{\Gamma
}\,C_{k+1}\, \,\,\,, \,\,\,\, \, \label{vnn0}
\end{eqnarray}
where\, $\,u_j={\bf v}_j -{\bf V}\,$. At boundaries of the forbidden
regions. i.e. at $\,|\rho_j|= r_0\,$, these equations should be
supplemented with boundary conditions (\ref{cr2})-(\ref{mr}).
Combining Eqs.\ref{vnn0} with (\ref{cr3}) and (\ref{cr2}) we come to
\begin{eqnarray}
\frac {\partial C_{k}}{\partial t}= -{\bf V}\!\cdot\! \frac {\partial
C_{k}}{\partial {\bf R}}+\sum_{j\,=1}^{k} \,\mathbf{L}_j^0\,C_{k}
 \,\, + \,\,\,\,\ \,\,\,\,\ \,\,\,\,\, \,\,\,\,\,\,
 \,\,\,\,\, \,\,\,\,\, \nonumber \\
+\,\frac {1}{\pi\lambda}\int d{\bf p} \int_{({\bf V}-{\bf
v})\cdot\Omega >\,0} d\Omega\, \,\, (({\bf V}-{\bf v})\cdot\Omega
)\,\times \,\,\,\,\,\, \,\,\,\,\,\,
\,\,\,\,\,\, \,\,\,\,\, \label{cbh} \\
\times\,\, [\,G_m({\bf p}^*)\, C_{k}({\bf P}^*) -\,G_m({\bf p})\,
C_{k}({\bf P})  \,+ \,\,\,\, \,\,\,\, \,\,\,\, \,\, \,\,\,\, \,
\nonumber \\ +\, C_{k+1}(\rho=r_0\Omega\,, {\bf P}^*,{\bf p}^* ) -
C_{k+1}(\rho=r_0\Omega\,, {\bf P},{\bf p}) \,] \,\,\nonumber
\end{eqnarray}
with\, $\,\rho =\rho_{k+1}\,$, $\,{\bf p} = {\bf p}_{k+1}\,$. $\,{\bf
v} = {\bf v}_{k+1} ={\bf p}_{k+1}/m\,$,\, and momenta $\,{\bf
P}^*\,$, $\,{\bf p}^*\,$ being related to $\,{\bf P}\,$, $\,{\bf
p}\,$\, in the same way as pre-collision momenta $\,{\bf
P}^{\,in}\,$, $\,{\bf p}^{\,in}\,$ in (\ref{mr}) are related to
post-collision ones, $\,{\bf P}^{\,out}\,$, $\,{\bf p}^{\,out}\,$,\,
and the integration involves pre-collision states only.

What is for the operators $\,\mathbf{L}_j^0\,$, in respect to CFs
they are defined as\, $\,\mathbf{L}_j^0 =-({\bf v}_j-{\bf V}) \cdot\
\partial /\partial \rho_j \,$\, at $\,|\rho_j|>r_0\,$\, and by
boundary conditions (\ref{cr2}) at $\,|\rho_j|=r_0\,$. Thus,
importantly, $\,\mathbf{L}_j^0\,$ are not mere translation operators:
at collision surfaces $\,|\rho_j|=r_0\,$ they act as creation
operators, creating $\,(k+1)$-order correlations from $\,k$-order
ones. Therefore factually $\,C_k\,$ remain coupled with both
$\,C_{k+1}\,$ and $\,C_{k-1}\,$,\, as in basic  Eqs.\ref{vn}. In
other words, due to conditions (\ref{cr2}), hierarchy (\ref{cbh})
keeps characteristic {\bf three-diagonal} structure of equations for
CFs\,!

Return from CFs back to DFs, according to the CFs definition
(\ref{cf}), transforms these equations to
\begin{eqnarray}
\frac {\partial F_{k}}{\partial t}= -{\bf V}\!\cdot\! \frac {\partial
F_{k}}{\partial {\bf R}}-\sum_{j\,=1}^{k}\,{\bf u}_j \cdot\! \frac
{\partial F_{k}}{\partial \rho_j} \, -\,n \,\widehat{\Gamma
}\,F_{k+1}\, \,\,\, \,\,\,\, \, \label{fnn0}
\end{eqnarray}
and conditions (\ref{cr2}) to
\begin{eqnarray}
F_{k}({\bf P}^{\,in},{\bf p}^{\,in}_j)\,= \,F_{k}({\bf
P}^{\,out},{\bf p}^{\,out}_j)\, \,\,\,\,\, \texttt{at} \,\,\,\,
\,\,\, |\rho_j|=r_0\,\,\, \,\,\, \,\,\, \,\,\, \label{crf}
\end{eqnarray}
along with (\ref{mr}). Their substitution into the ``collision
integral' $\,\widehat{\Gamma }\,F_{k+1}\,$ gives, similarly to
(\ref{cbh}),
\begin{eqnarray}
\frac {\partial F_{k}}{\partial t}= -{\bf V}\!\cdot\! \frac {\partial
F_{k}}{\partial {\bf R}}\,+\, \sum_{j\,=1}^{k}\,\mathbf{L}_j^0\,F_{k}
\,\, + \,\,\,\,\ \,\,\,\,\
\,\,\,\,\, \,\,\,\,\,\, \,\,\,\,\, \,\,\,\,\, \nonumber \\
+\,\frac {1}{\pi\lambda}\int d{\bf p} \int_{({\bf V}-{\bf
v})\cdot\Omega
>\,0} d\Omega\, \,\, (({\bf V}-{\bf v})\cdot\Omega )\,\times \,\,\,\,\,\,
\,\,\,\,\,\,
\,\,\,\,\,\, \,\,\,\,\, \label{fbh} \\
\times\,\, [\, F_{k+1}(\rho=r_0\Omega\,, {\bf P}^*,{\bf p}^* )
-F_{k+1}(\rho=r_0\Omega\,, {\bf P},{\bf p}) \,]\,\,  \,\, \nonumber
\end{eqnarray}
Here again\, $\,\mathbf{L}_j^0 =-({\bf v}_j-{\bf V}) \cdot\
\partial /\partial \rho_j \,$\, at $\,|\rho_j|>r_0\,$\,
but at $\,|\rho_j|=r_0\,$\, action of the operator
$\,\mathbf{L}_j^0\,$ onto DFs is defined by boundary conditions
(\ref{crf}).

\section{The Boltzmann-Grad limit and ``mathematical non-physics''}

Equations (\ref{vnn0}) as combined with boundary conditions
(\ref{cr2})-(\ref{mr}) at collision surfaces $\,|\rho_j|=r_0\,$ (and
conditions (\ref{icv}) at infinity) or, equivalently, equations
(\ref{fnn0}) combined with (\ref{crf}) and (\ref{mr}) represent
direct analogue of so-called  ``hard-sphere BBGKY hierarchy'' (see
e.g. Eq.2.2 from \cite{vblls} or Eq.4.1 from \cite{sp1}).

Importantly, the term ``hard-sphere BBGKY hierarchy'' is adequate on
the understanding only that\, when the boundary conditions are
substituted into $\,F_{k+1}\,$ inside the ``collision integral'' in
equation for $\,F_{k}\,$\, then {\bf simultaneously and necessarily}
they are satisfied by $\,F_{k+1}\,$ in the next equation for
$\,F_{k+1}\,$ itself (i.e. included into definition of the operators
$\,\mathbf{L}_j^0\,$ as above). Otherwise one makes some
``mathematical non-physics'', since resulting equations will be
non-derivable from Liouville equation for an (infinitely) many hard
sphere system\,!

In opposite, fulfilment of the mentioned requirement guarantees
observance of the virial relations (see Appendix) and consequently
belonging of resulting equations to the class of BBGKY hierarchies
(since virial relations do follow already from most general
properties of many-particle dynamics and Liouville equations
\cite{tmf,artmf,pro}).

In view of these facts, it seems impossible to accept the old idea
that under the Boltzmann-Grad limit (BGL) the ``hard-sphere BBGKY
hierarchy'' is equivalent to so-called ``hard-sphere Boltzmann
hierarchy'' (see e.g. \cite{lan,vblls,sp1,gp2}). For our system it
looks as
\begin{eqnarray}
\frac {\partial F_{k}}{\partial t}= -{\bf V}\!\cdot\! \frac {\partial
F_{k}}{\partial {\bf R}}-\sum_{j\,=1}^{k}\,({\bf v}_j -{\bf V})
\cdot\! \frac {\partial F_{k}}{\partial \rho_j} \,\, + \,\,\,\,\
\,\,\,\,\
\,\,\,\,\, \,\,\,\,\,\, \,\,\,\,\, \,\,\,\,\, \nonumber \\
+\,\frac {1}{\pi\lambda}\int d{\bf p} \int_{({\bf V}-{\bf
v})\cdot\Omega >\,0} d\Omega\,\,\, (({\bf V}-{\bf v})\cdot\Omega )
\,\times \,\,\,\,\,\, \,\,\,\,\,\,
\,\,\,\,\,\, \,\, \label{fbh0} \\
\times\,[\, F_{k+1}(\rho=0\,, {\bf P}^*,{\bf p}^* ) -
F_{k+1}(\rho=0\,, {\bf P},{\bf p}) \,]\,\,\,, \nonumber
\end{eqnarray}
where there are no forbidden regions, that is $\,\rho_j\,$ can take
arbitrary values from the whole $\,\mathbb{R}^3\,$, and the boundary
conditions (\ref{crf}) at $\,|\rho_j|\rightarrow r_0 \rightarrow 0\,$
are thrown, that is operators $\,\mathbf{L}_j^0\,$ are replaced by
trivial translation operators,\, $\,-({\bf v}_j -{\bf V}) \cdot\,
\partial /\partial \rho_j\,$.

This means that in corresponding equations for CFs any $\,C_{k} \,$
is now connected to $\,C_{k+1}\,$ only, and not to $\,C_{k-1}\,$.
Thus, these equations form {\bf two-diagonal} hierarchy {\bf
qualitatively different} from the original one\,!

As the result of such frivolity, these equations allow for factored
solution with ``propagation of chaos'', when\, $\,F_k=F_0
\prod_{j=1}^k G_m({\bf p}_j)\,$,\, $\,C_k= 0\,$ at $\,k>0\,$.\, and
$\,F_0=C_0\,$ undergoes the ``Boltzmann-Lorentz equation''.

Evidently, this ``Boltzmann hierarchy'' contraries to the above
emphasized requirement:\, it uses the boundary conditions to write
``collision integrals'' but neglects the same conditions  in higher
equations what determine the integrands (as if the latter took
particles from a ``parallel world'')\,! As the consequence, the
Boltzmann hierarchy does not satisfy the virial relations. Its wrong
is clear already from observation that it results when one first
damages Eqs.\ref{fbh} by replacing $\,\mathbf{L}_j^0\,$ with\,
$\,-({\bf v}_j -{\bf V}) \cdot\,
\partial /\partial \rho_j\,$\, and only after that goes to
$ \,r_0=0\,$\, \cite{comm3}. Therefore, it is not surprising that the
famous Lanford's attempt \cite{lan} to prove the mentioned idea was
in fact unsuccessful \cite{i2,comm4}.

Possible formal cause of this nonsuccess is very simple: the smaller
is $\,r_0\,$ the less smooth functions of $\,\rho_j\,$ are all the
CFs \cite{pro,mpa3}, since conditions (\ref{crf}) ensure continuity
and smoothness of the probability measures $\,F_k\,$ along phase
trajectories only but not in perpendicular dimensions.

Indeed, let $\,{\bf u}_j\cdot \rho_j >0\,$, that is BP and an atom
scatter one from another. If at that $\,r_0\ll
|\rho_j|\lesssim\lambda\,$\, and additionally\, $\,|{\bf u}_j \times
\rho_j |/|{\bf u}_j | < r_0\,$\, (where $\,\times\,$ denotes vector
product), i.e. vectors $\,{\bf u}_j\,$ and $\,\rho_j\,$ are nearly
parallel, then the particles form an {\it out\,}-state which (almost
certainly) arose from their recent collision (especially if
$\,|\rho_j|\ll \lambda\,$).According to (\ref{cr2}), post-collision
correlations of any order between such particles inevitably take
place, and by order of magnitude all they are equal to $\,C_0=F_0\,$.
With time, as was explained in Sec.3, also quite similar
pre-collision correlations do arise, at $\,{\bf u}_j\cdot \rho_j
<0\,$ and again at $\,|{\bf u}_j \times \rho_j |/|{\bf u}_j | <
r_0\,$, since statistical picture acquires more and more
time-reversal symmetry.

At the same time, at $\,|{\bf u}_j \times \rho_j |/|{\bf u}_j | >
r_0\,$ there are almost no reasons for correlations. Hence, in
directions perpendicular to $\,{\bf u}_j\,$ all CFs become more and
more sharp functions of $\,\rho_j\,$. At that. the integral value of
correlation per one atom goes to zero $\,\propto \pi r_0^2\lambda
=1/n\,$. But this has no significance since correlations concentrate
just where they are most effective. This is confirmed by the virial
relations (\ref{intc}). The latter show also that really significant
correlational characteristics are $\,n^k C_k\,$ which turn under BGL
into singular generalized functions.

\section{Conclusion}

We have seen that the Boltzmann-Grad limit does not lead to any
essential simplifications (and in this sense it does not exist). In
particular, it does not lead to the mythologic Botzmannian kinetics
with Boltzmann equation, Boltzmann-Lorentz equation, etc.), except
non-interesting spatially homogeneous case \cite{kac}. Physical and
statistical reasons of this pleasant fact already were over and over
again explained in \cite{i1,bbgky,i2} and earlier
\cite{pr1,bk1,bk2,pr2,bk3,pjtf} and later
\cite{tmf,pro,mpa3,p12,ig,i3,i4}. The fact is pleasant because it
shows once again that real many-particle dynamical chaos does not
degenerates into miserable stochastics.

\appendix*
\section{\,\,Virial relations for hard-sphere system}

Integration of $\,(k+1)$-th of Eqs.\ref{vnn0} over $\,{\bf
p}_{k+1}\,$ and $\,\rho_{k+1}\,$ at $\,|\rho_{k+1}|>r_0\,$\, yields
for $\,\widetilde{C}_k=\int_{k+1} C_{k+1}\,$, in view of (\ref{icv}),
equations
\[
\frac {\partial \widetilde{C}_{k}}{\partial t}= -{\bf V}\!\cdot\!
\frac {\partial \widetilde{C}_{k}}{\partial {\bf
R}}-\sum_{j\,=1}^{k}\,{\bf u}_j \cdot\! \frac {\partial
\widetilde{C}_{k}}{\partial \rho_j} \,-\,\widehat{\Gamma }\,C_{k+1}
-\,n \,\widehat{\Gamma }\,\widetilde{C}_{k+1}\, \,
\]
At the same time, differentiation of Eqs.\ref{vnn0} in respect to the
density produces for $\,\overline{C}_k=\partial C_{k}/\partial n\,$\,
equations
\[ \frac {\partial
\overline{C}_{k}}{\partial t}= -{\bf V}\!\cdot\! \frac {\partial
\overline{C}_{k}}{\partial {\bf R}}-\sum_{j\,=1}^{k}\,{\bf u}_j
\cdot\! \frac {\partial \overline{C}_{k}}{\partial \rho_j}
\,-\,\widehat{\Gamma }\,C_{k+1} -\,n \,\widehat{\Gamma
}\,\overline{C}_{k+1}\, \,
\]
which are identical to the previous ones. Since, according to
(\ref{icv}), initial conditions for $\,\overline{C}_k\,$ and
$\,\widetilde{C}_k\,$ also are identical (all are zeros), we come to
equalities $\,\overline{C}_k =\widetilde{C}_k\,$, that is to the
virial relations (\ref{me}),\, $\,\partial C_k/\partial
n\,=\int_{k+1} C_{k+1}\,$.

Notice that this consideration is valid regardless of concrete form
of boundary conditions at $\,|\rho_{j}|=r_0\,$. Therefore, virial
relations hold under replacement of the hard-sphere ``mirror
reflection'' conditions (\ref{mr}) by some other rule.
Correspondingly, equations (\ref{vnn0}) and (\ref{fnn0}) can be
extended to collision operators\, $\,\widehat{\Gamma }\,$\, with
other ``scattering laws'' than (\ref{mr}).



\begin{thebibliography}{41}
\bibitem{ex}
 \,Similarly, in the human life, if somebody is
so much free that forgets his past expenses, then he can not plan
``mean expenses per unit time''!
\bibitem{com}
\,In the low-density gas limit these fluctuations become even more
(maximally) pronounced since get rid of damping by collective
(hydrodynamical) interactions.\, As these fluctuations have no
specific reasons, one can say that merely they manifest
non-ergodicity of kinetic behavior of many-particle dynamical systems
\cite{tmf,artmf,i1,i2}.
\bibitem{rev}
\,Unfortunately, so trivial things are beyond field of vision of
reviewers who easy write such nonsenses as: ``statistical
correlations at low density have little influence on the motion of
particles'' (how do you like it!). In fact. of course, statistical
correlations never influence on some real processes, since reside in
theory. But there (at arbitrary density!) they are the only
representatives of any orocesses including particles' interactions
and collisions.
\bibitem{comm2}
\,In my previous papers thece functions were designated as $\,V_k\,$
in place of $\,C_k\,$.
\bibitem{comm3}
\,Apparently, just such damaged equations are taken in mind as
``hard-sphere BBGKY hierarchy'' by amateurs of ``Boltzmann
hierarchy''.
\bibitem{comm4}
\,May be, more careful and fine considerations like e.g. \cite{sp2}
will make this comprehensible to all.
\end{thebibliography}
\end{document}